\documentclass[prl,aps,twocolumn,showpacs]{revtex4}
\usepackage{graphicx}
\usepackage{dcolumn}
\newcommand{\nn}{\nonumber}

\begin{document}

\title{Direct Imaging of Spatially Modulated Superfluid Phases in Atomic Fermion Systems}

\author{T. Mizushima}
\affiliation{Department of Physics, Okayama University,
Okayama 700-8530, Japan}
\author{K. Machida}
\affiliation{Department of Physics, Okayama University,
Okayama 700-8530, Japan}
\author{M. Ichioka}
\affiliation{Department of Physics, Okayama University,
Okayama 700-8530, Japan}
\date{\today}

\begin{abstract}

It is proposed that the spatially modulated superfluid phase, or
the Fulde-Ferrell-Larkin-Ovchinnikov (FFLO) state could be observed
in resonant Fermion atomic condensates which are realized recently.
We examine optimal experimental setups to achieve it by solving
Bogoliubov-de Gennes equation both for idealized one-dimensional
and realistic three-dimensional cases. The spontaneous modulation
of this superfluid is shown to be directly imaged as the density profiles 
either by optical absorption or by Stern-Gerlach experiments.

\end{abstract}

\pacs{03.75.Ss, 03.75.Hh, 74.20.Fg}

\maketitle

A keen interest focuses on resonance Fermion condensations by using ultracold atomic $^6$Li or
$^{40}$K clouds. It is now agreed that the BCS superfluid state has been achieved by various experimental 
groups \cite{jochim,greiner,zwierlein,regal,zwierlein2,kinast,bourdel}, 
where magnetic field Feshbach resonance is utilized to finely control the atomic interaction 
between strong repulsive and attractive regions. 
There are not only active experimental and theoretical \cite{holland,ohashi}
investigations on this BCS-BEC cross-over phenomenon, but also pioneering  works 
on exploring the BCS pairing state, including  spectroscopic experiments \cite{chin,greiner2} 
to extract the energy gap value
as functions of temperatures and the coupling constant \cite{kinnunen,buchler}.

In the current experiments, the prepared Fermion numbers of two species, where the hyperfine states 
$| F=\frac{9}{2}, m_F=-\frac{9}{2}\rangle$ and $|F=\frac{9}{2}, m_F=-\frac{7}{2}\rangle$ are used
in the $^{40}$K case \cite{jochim}, are equal to maximize the transition temperature.
Here by intentionally making these numbers unequal, we could achieve a 
spatially modulated BCS state, namely so-called Fulde-Ferrell-Larkin-Ovchinnikov (FFLO) state \cite{ff,lo}.

This interesting state has been much sought out for long time in vain in a superconductor. One of the main difficulty to stabilize it lies in the fact that applying field to create population difference of up and down electrons inevitably induces diamagnetic current which acts a Cooper pair breaker. An exceptional case is ferromagnetic superconductors whose internal ferromagnetic molecular field does not induce it. Even under this favorable situation any firm conclusion has not been reached for ternary compounds ErRh$_4$B$_4$ and HoMo$_6$S$_8$\cite{machida,machida2}. One of the difficulties to observe it is the absence of an experimental method to see the modulation. The FFLO state is also discussed in a new heavy compound CeCoIn$_5$ \cite{kakuyanagi} or even in high energy community in connection with neutron stars \cite{casalbuoni}.

The present neutral Fermion systems are ideal to pursue this realization of the FFLO state because: (1) There is no pair breaking mechanisms such as the diamagnetic current or impurities which weaken the stability of this state. (2) As is shown below, fine adjustment on the population difference of two species, say, up and down Fermions, and attractive interactions is possible for the system to be in most favorable and easily observable condition for the FFLO state.

The purposes of the present paper are to analyze the most suitable condition to observe the FFLO state by maximally utilizing the limited experimental resources for the present neutral atomic vapors and also to provide the experimental signatures of the FFLO state within the currently available experimental techniques.
The arrangement of this paper is following: After introducing 
the Bogoliubov-de Gennes formalism which allows us to describe the spatially modulated phase, we provide briefly known analytic solution for the FFLO state 
of the quasi-one-dimensional system \cite{machida2}, which is useful to grasp the general perspective of the present problem. Then, we go on showing a numerical calculation for the more realistic situation where the atomic cloud is confined by a three-dimensional harmonic trap.

We start out with the pairing Hamiltonian
\begin{eqnarray}
&&
\mathcal{H} = \int d{\bf r} 
 \left[
	\sum _{\sigma=\uparrow, \downarrow} \hat{\psi}^{\dag}_{\sigma}({\bf r}) 
	\left\{ - \frac{\hbar^2}{2m} \nabla^2 +V({\bf r}) - \mu _{\sigma} \right\} \hat{\psi} _{\sigma}({\bf r}) 
\right. \nn \\
&& \hspace{2cm} \left.
	+ \frac{g}{2} \sum _{\sigma,\sigma'=\uparrow, \downarrow} 
	\hat{\psi}^{\dag}_{\sigma}({\bf r}) \hat{\psi}^{\dag}_{\sigma'}({\bf r})
	\hat{\psi}_{\sigma'}({\bf r}) \hat{\psi}_{\sigma}({\bf r})
\right],
\end{eqnarray}
with $g=\frac{4\pi \hbar^2 a}{m}$ being the attractive interaction.  $a(<0)$ is the $s$-wave scattering length. An axis-symmetric trap potential $V({\bf r}) = \frac{1}{2}m (\omega^2_{r} r^2 + \omega^2_z z^2)$ is assumed. Here, to prepare the unequal population in two species $\sigma = \uparrow, \downarrow$, the following chemical potential is used: $\mu _{\sigma} = \mu + (\hat{\sigma}_3)_{\sigma,\sigma} \delta\mu$ with the Pauli matrix $\hat{\sigma}_3$. The Fermion operators of the two species are denoted by $\hat{\psi}^{\dag}_{\sigma}({\bf r})$ and $\hat{\psi}_{\sigma}({\bf r})$. The Bogoliubov transformation with the mean-field approximation is employed to diagonalize the Hamiltonian, that is, $\hat{\psi}_{\uparrow} = \sum _{\bf q}[u_{\bf q}\eta _{{\bf q},\uparrow} - v^{\ast}_{\bf q}\eta^{\dag}_{{\bf q},\downarrow}]$ and $\hat{\psi}^{\dag}_{\downarrow} = \sum _{\bf q}[v_{\bf q}\eta _{{\bf q},\uparrow} - u^{\ast}_{\bf q}\eta^{\dag}_{{\bf q},\downarrow}]$ within the creation and annihilation operators $\eta _{{\bf q},\sigma}$ and $\eta ^{\dag}_{{\bf q},\sigma}$. To this end, one can lead to the Bogoliubov-de Gennes (BdG) equation:
\begin{eqnarray}
\left[ 
	\begin{array}{cc}
		\mathcal{K}_{\uparrow} - \mu _{\uparrow} & \Delta ({\bf r}) \\
		\Delta ^{\ast} ({\bf r}) & - \mathcal{K}^{\ast}_{\downarrow} + \mu _{\downarrow}
	\end{array}
\right] 
\left[ 
	\begin{array}{c} u_{\bf q}({\bf r}) \\ v_{\bf q}({\bf r}) \end{array}
\right] =  \varepsilon _{\bf q}
\left[ 
	\begin{array}{c} u_{\bf q}({\bf r}) \\ v_{\bf q}({\bf r}) \end{array}
\right],
\label{eq:bdg}
\end{eqnarray}
where $\mathcal{K}_{\sigma} = - \frac{\hbar^2}{2m}\nabla^2 +V({\bf r}) + g\rho _{\sigma} ({\bf r})$ with the self-consistent equation
\begin{eqnarray}
\Delta({\bf r}) = g_{\rm eff} \sum_{\bf q} u_{\bf q}({\bf r}) v^{\ast}_{\bf q}({\bf r}) f(\varepsilon _{\bf q}). 
\label{eq:gap}
\end{eqnarray} 
The order parameter $\Delta ({\bf r})= g _{\rm eff} \langle \hat{\psi}_{\downarrow}({\bf r}) \hat{\psi}_{\uparrow} ({\bf r}) \rangle$ and the particle density of each component $\rho _{\uparrow}({\bf r}) = \sum _{\bf q} |u_{\bf q}({\bf r})|^2 f(\varepsilon _{\bf q})$, and $\rho _{\downarrow}({\bf r}) = \sum _{\bf q} |v_{\bf q}({\bf r})|^2 [1 - f(\varepsilon _{\bf q}) ]$ where $f(\varepsilon _{\bf q}) = 1/(e^{\varepsilon _{\bf q}/k_BT} + 1)$ is the Fermi-distribution function. Since the finite shift of the chemical potential $\hat{\sigma}_3\delta\mu$ causes a distinction of the mean-field potential for two species, the particle-hole symmetry is broken \cite{memo}. Therefore, the sum in Eq.~(\ref{eq:gap}), is done for all the eigenstates with both positive and negative eigenenergies. When calculating $\Delta ({\bf r})$, we have used a regularized coupling constant $g_{\rm eff}$ to avoid  the ultraviolet divergence \cite{bulgac}.

In the one spatial dimension without a trap, which is approximate for a long cigar-shaped gas  $\omega _z / \omega _r \rightarrow 0$, 
one admits an exact analytical solution \cite{machida2}, which has the form
\begin{eqnarray} 
\Delta (z) = \Delta _1 k _1 {\rm sn}(\Delta _1 z, k_1). 
\end{eqnarray}
Here, ${\rm sn}(z,k)$ is the Jacobi elliptic function with the modulus $k$ ($k_1$ complementary modulus). 
$\Delta_1$ is the order parameter to be determined self-consistently.
According to this solution at zero-temperature, 
the FFLO state is energetically stable for the relative population difference 
\begin{eqnarray}
\delta n \equiv \frac{|n _{\uparrow}-n _{\downarrow}|}{n} \ge \frac{1}{\pi}\frac{\Delta _0}{\varepsilon _F},
\end{eqnarray}
where the particle number of each component $n_{\sigma} = \int d{\bf r} \rho _{\sigma} ({\bf r})$ and the total particle number $n=\sum _{\sigma} n_{\sigma}$.
$\varepsilon _F$ is the Fermi energy and $\Delta _0$ is the amplitude of 
the order parameter of a uniform BCS state at zero-temperature. 
Beyond the above critical population difference the uniform BCS state changes into the modulated FFLO state. 
Since $\Delta _0/\varepsilon _F = 0.2 \sim 0.4$ in the present experiments, 
the critical population difference should be of an order of $10\% \sim 20 \%$ difference.

This modulated phase accompanies the spin variation with half of the fundamental modulation periodicity.
The excess density of the up-spin Fermions ($n_{\uparrow}>n_{\downarrow}$) periodically accumulates 
at the zeros of $\Delta(z)$. The accumulation becomes sharp in space when either approaching 
the critical population difference or stronger coupling region.
It is illustrated in Figs. 11 and 12 of Ref.~\cite{machida2} where $\Delta(z)$ and the
associated spin modulation are seen for two typical $\delta n$ values, 
one is near the critical boundary and the other far from it.

Since each atomic species can be probed optically because of different absorption frequencies, 
we can image each density separately in space.
Thus it is crucial to appropriately choose the two parameters; 
$|\delta n| \equiv |n_{\uparrow} - n_{\downarrow}|/n$ and $\Delta _0/\varepsilon _F$ 
in order to obtain a sharp absorption image. 
Another way to detect this spin modulation is to perform Stern-Gerlach experiments, 
revealing the different particle density distributions for spin-up and spin-down Fermions.

As for the modulation periodicity $\lambda$ relative to the mean particle distance $L=1/k_F$, 
\begin{eqnarray}
\frac{\lambda}{L} = \frac{\varepsilon _F}{\Delta _0}\frac{kK(k)}{\pi},
\end{eqnarray}
where $K(k)$ is the complete elliptic integral of the first kind.
As a rough estimate we find $\lambda = 3\mu {\rm m} \sim 30 \mu {\rm m}$ for 
$\Delta / \varepsilon _F = 0.1 \sim 0.01$ and $n \sim 10^{20} {\rm m}^{-3}$.

After having finished the order of magnitude estimates for various quantities based on the analytic solution, we present a more realistic calculation by numerically solving the BdG equation (\ref{eq:bdg}) self-consistently for help to design an experiment: We consider the axis-symmetric harmonic trap potential with $\omega _r /2\pi \!=\! 1.67 \omega _z/2\pi \!=\! 1000$Hz and the $^6$Li atoms with the chemical potential $\mu \!=\! 12.5 \hbar\omega _r$, corresponding to $n\sim 1100$ atoms. Throughout this paper, $\Delta _0/\varepsilon _F \!=\! 0.35$ and $k_BT/\hbar\omega _r \!=\! 0.05$ are fixed.

\begin{figure}[t]
\includegraphics[width=7.5cm]{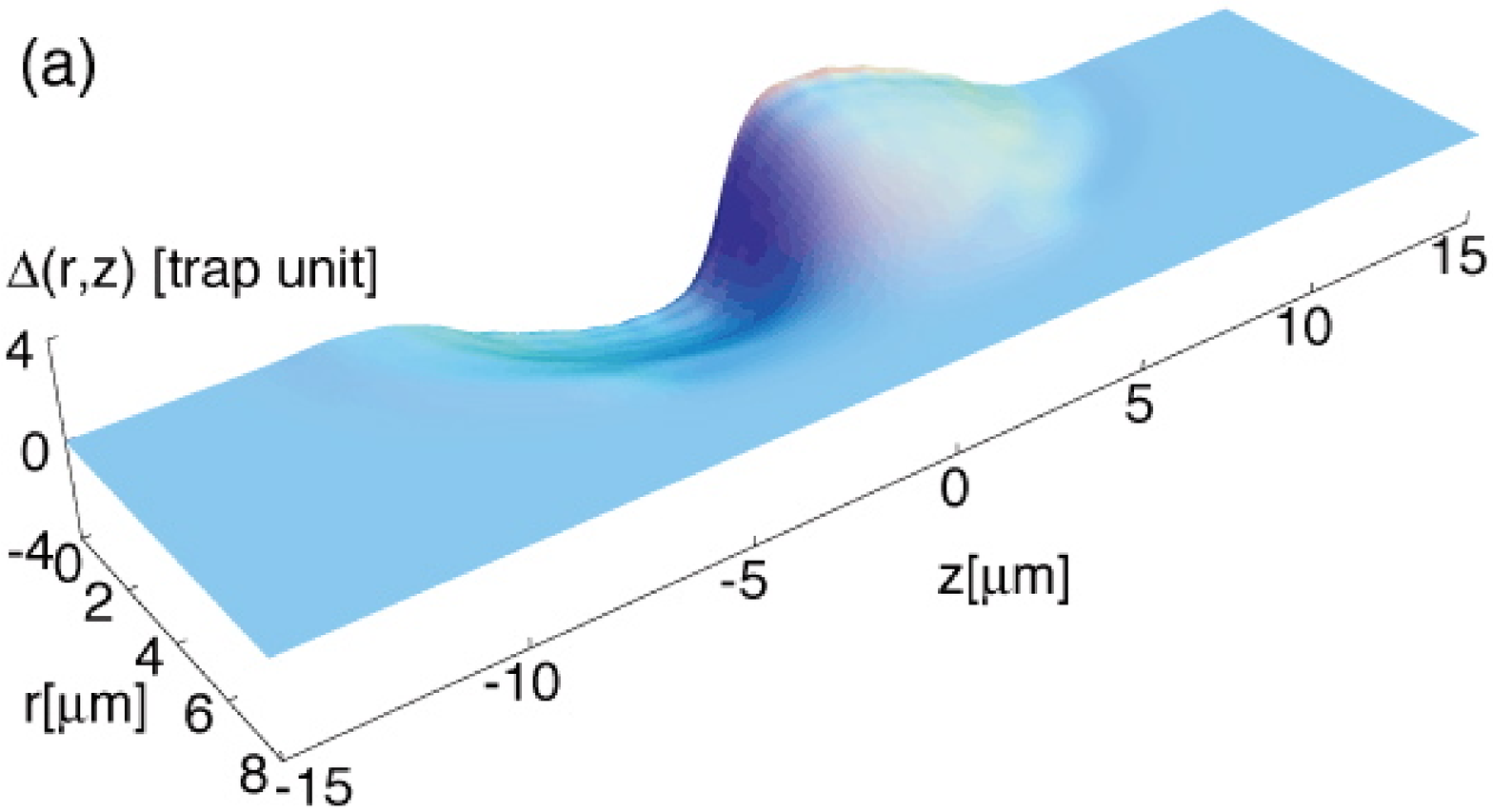} \\
\includegraphics[width=7.5cm]{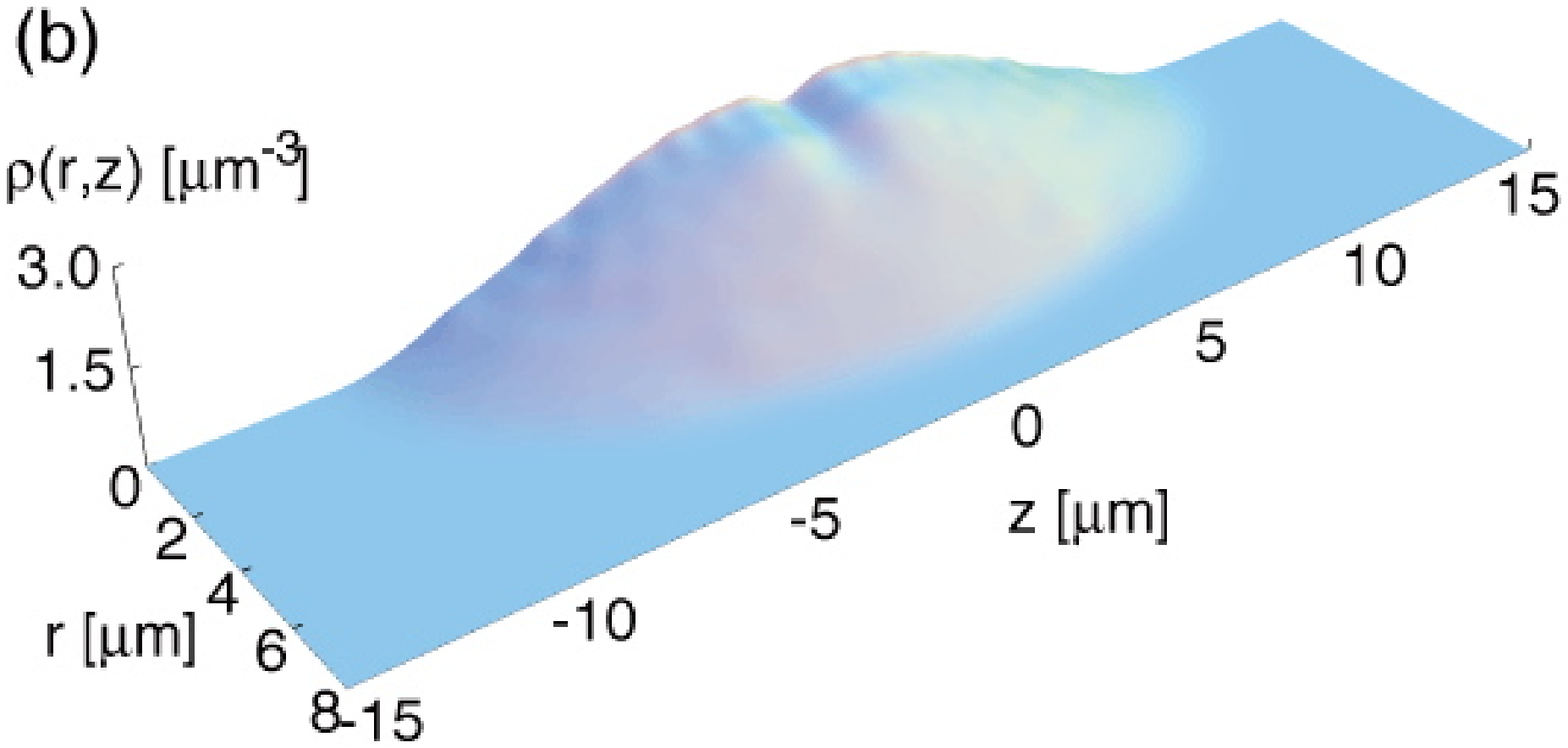} \\
\includegraphics[width=8cm]{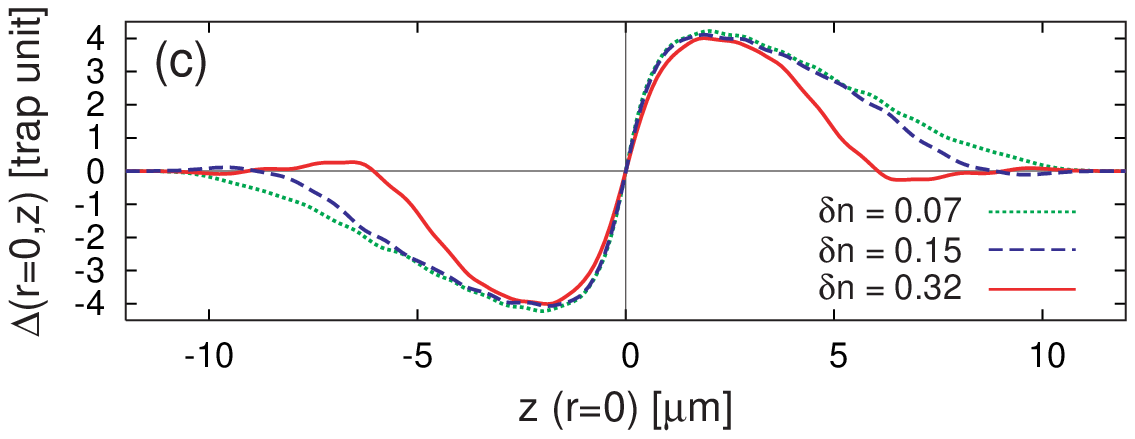} 
\caption{
(a) The spatial profiles of the pairing field $\Delta (r,z)$ and (b) the total density $\rho (r,z)$ 
for the population difference $\delta n=0.15$ and $\Delta _0 /\varepsilon _F = 0.35$. 
(c) The cross-section of $\Delta (r,z)$ at $r=0$ is shown for various population differences $\delta n$.
}
\label{fig:gap}
\end{figure}

The results are displayed in Fig.~\ref{fig:gap} where the spatial variation of the pairing field $\Delta (r,z)$ and the total density profile $\rho(r,z)$ are shown. $\Delta (r,z)$ changes its sign at the plane $z=0$ and also near the edges of the pairing field (See the red curve in Fig.~\ref{fig:gap}(c)). Since the system has an elongated ellipsoidal shape, $\Delta (r,z)$ is never a simple sinusoidal form. This is quite different from the above one-dimensional
case where the total density is assumed to be uniform. In the present situation
the total density is non-uniform because of the confinement. 
The order parameter is inevitably inhomogeneous intrinsically.

As is seen from Fig.~\ref{fig:gap}(b) the total density along the $z$-axis 
is similar to that expected for Thomas-Fermi approximation for the normal Fermions except for a small feature, 
a dip at $z=0$. This is related to the local suppression of $\Delta({\bf r})$.

In Fig.~\ref{fig:gap}(c) we show the cross-section $\Delta(r=0,z)$ at the $r=0$ axis. As $\delta n$ becomes large, 
the modulation period tends to become short, thus the extra zeros appear near the edges. 
It is also noted from this that for $\delta n=0.32$
the modulation period becomes shorter towards the edges where the local density tends to be dilute.

\begin{figure}[t]
\includegraphics[width=7.5cm]{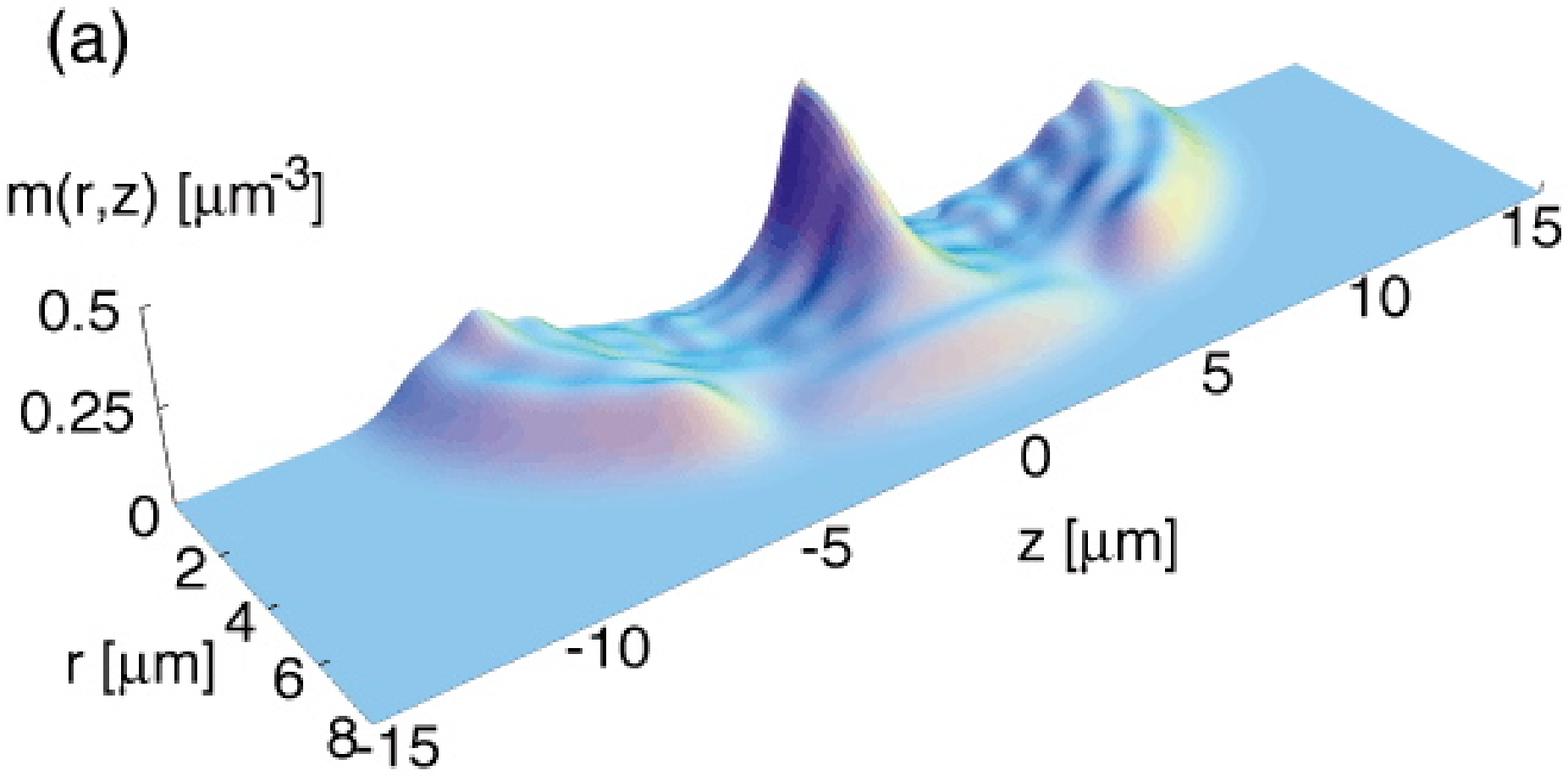} \\
\includegraphics[width=8cm]{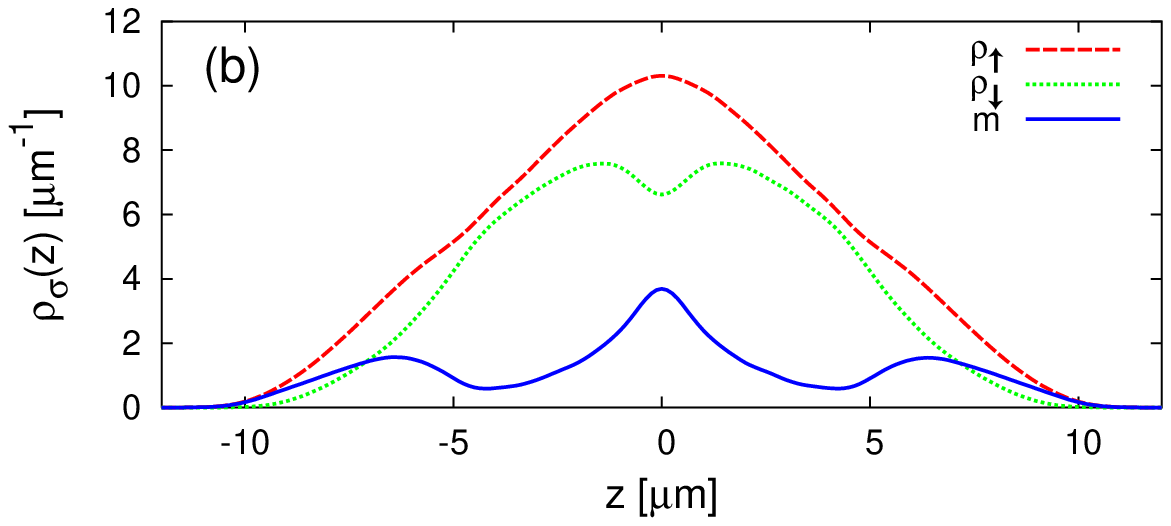} \\
\includegraphics[width=8cm]{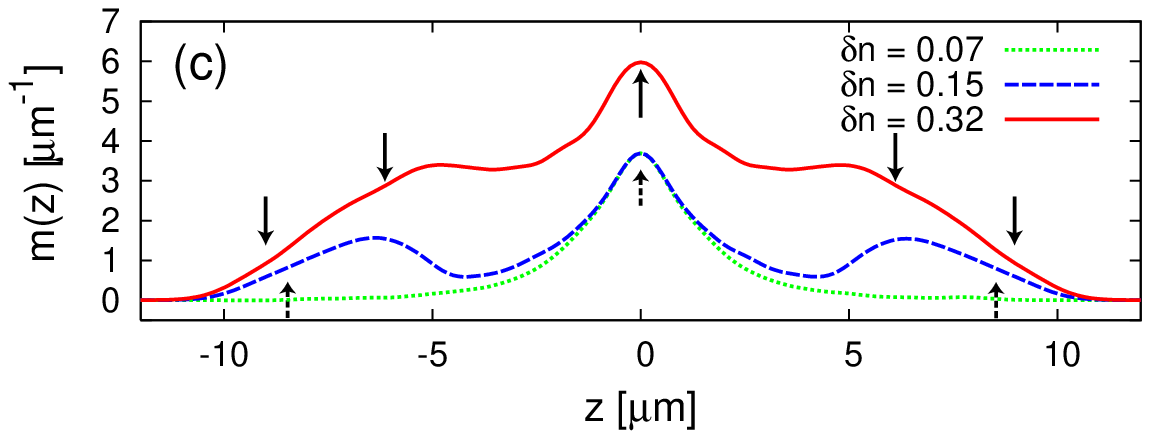} 
\caption{
(a) Stereographic view of the magnetization $m(r,z)$, and (b) its columnar
density profile along the $z$-axis for $\delta n=0.15$.
(c) The columnar magnetization profiles $m(z)$ for various $\delta n= 0.07,0.15,0.32$. 
The arrows indicate the positions of the zeros of $\Delta(r=0,z)$, which is displayed in Fig.~\ref{fig:gap}.
The number of the zeros is 1 ($\delta n = 0.07$), 3 ($\delta n = 0.15$), and 5 ($\delta n = 0.32$).
}
\label{fig:density}
\end{figure}

In Fig.~\ref{fig:density} we display our most important results, which are directly 
probed experimentally. The local magnetization  $m({\bf r})$ defined by 
$m({\bf r})=\rho _{\uparrow}({\bf r})- \rho _{\downarrow}({\bf r})$, 
which is the local population difference of the two species, is shown in Fig. 2(a). 
It is seen from this that $m(\bf r)$ exhibits peaks at the central plane $z=0$, 
where a prominent peak is observed at $r=0$, and also a small peak at the edge. 
Since we are treating a three dimensional object with an almost spherical shape, 
it is essential for experiments that the physical quantities integrated along the radial direction, 
namely the columnar densities should exhibit a prominent feature to probe it.
In Fig.~\ref{fig:density}(b) we display the columnar density  
$\rho _{\sigma}(z)\equiv 2\pi\int dr r\rho _{\sigma}(r,z)$ 
and $m(z)= \rho _{\uparrow}(z) - \rho _{\downarrow}(z)$,
both of which are directly observable. Because the two species have different
magnetic moments we can spatially separate the two species by passing it through a field gradient. 
In this Stern-Gerlach experiment, which is performed in spinor BEC \cite{stenger}, 
$\rho _{\uparrow}(z)$ and $\rho _{\downarrow}(z)$ in Fig. 2(b) are directly imaged.
Another way to reveal these modulations is to utilize the fact that each species has different clock transition frequency. By taking an absorption image with a particular frequency tuned we can selectively probe these modulations.

It is seen from Fig. 2(c) that depending on the population difference under a fixed $\Delta _0/\varepsilon _F$, the magnetization $m(z)$ shows quite different spatial structures: As $\delta n$ increases, the single peak structure for $\delta n=0.07$ changes into multiple peak ones, corresponding to the increment of the zeros of $\Delta(r,z)$ 
(also see Fig.~\ref{fig:gap}(c)).

\begin{figure}[t]
\includegraphics[width=8cm]{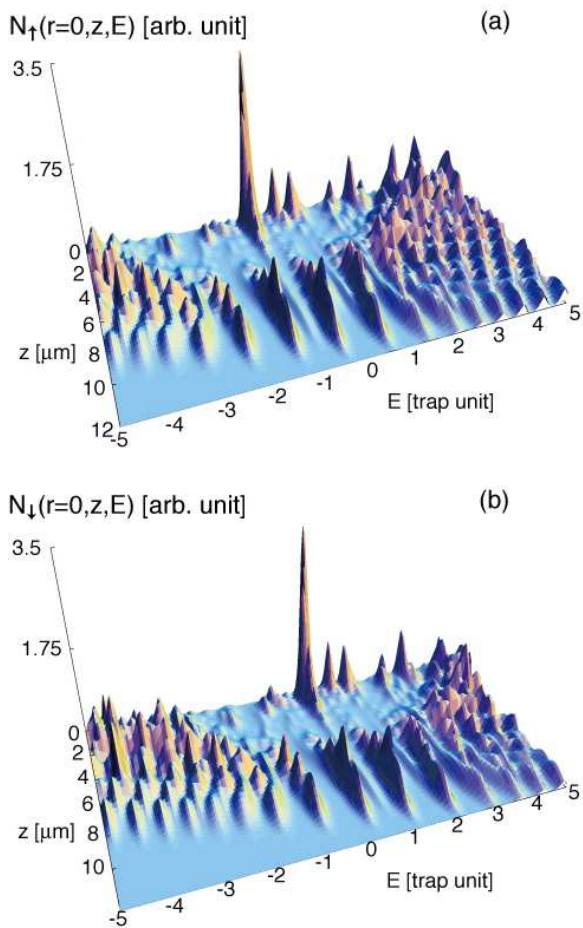} 
\caption{
(a) The local density of states profiles $N_{\uparrow}(r=0,z,E)$ and (b) $N_{\downarrow}(r=0,z,E)$ at $r = 0$
for the population difference $\delta n=0.15$ and $\Delta _0 /\varepsilon _F = 0.35$.
The large empty region in the center in both figures corresponds to the energy gap. 
}
\label{fig:ldos}
\end{figure}

It is instructive to see the gap structure in the FFLO state because the Fermionic excitation spectrum is distinctively altered from the uniform BCS state. In Fig.~\ref{fig:ldos} the local density
of states (LDOS) for up and down spins defined by
\begin{eqnarray}
&& N_{\uparrow}({\bf r},E) = \sum _{\bf q} |u_{\bf q}({\bf r})|^2 \delta(E - \varepsilon _{\bf q}),  \nn \\
&& N_{\downarrow}({\bf r},E) = \sum _{\bf q} |v_{\bf q}({\bf r})|^2 \delta(E+\varepsilon _{\bf q}),
\end{eqnarray}
are shown.
It is seen from Fig.3 that: (i) Away from the center at $r=0$ and $z=0$, 
the energy gap becomes narrow as a function of $z$, 
corresponding to the low-energy quasiparticle excitations bounded near the surface of the cloud \cite{baranov}.
(ii) The localized LDOS appears near the Fermi level at $z=0$ and also near the edges of the cloud $z \sim 8 \mu$m. In the spin-up (spin-down) case of Fig.~\ref{fig:ldos}(a) ((b)) the large mid-gap state at $z=0$ situating below (above) the Fermi level is filled (emptied), meaning physically that the excess spin-up Fermions are accommodated by keeping the overall bulk gap structure intact.
(iii) These mid-gap states at $z=0$ correspond to the positions of the zeros of $\Delta(r,z)$ shown in Fig.1. 

We can observe these mid-gap states by using stimulated Raman transition as an extra satellite.
This technique is employed by Chin {\it et al.} \cite{chin}, who identify the energy gap in Fermi condensates.
Note that this spectroscopy is most useful for a cigar shape cloud to separate out the mid-gap and surface excitations.

Finally we should mention a work by Combescot \cite{combescot} who examines the stability problem of neutral atomic Fermions with different Fermi radii under the absent of a trap. Yet another superfluid phase of an anisotropic gap is predicted. The precise boundary between FFLO and this phase remains to be worked out in a realistic situation.

We also mention that the present spatially modulated in FFLO shares a common physics of the so-called stripes paradigm, which includes spin density waves in Cr \cite{machida3}, spin-Peierls system \cite{fujita}, and high-$T_c$ cuprates \cite{machida4}. In these systems the order parameter of the staggered moment is commonly characterized by the spatial modulation with sign change. The node of the order parameter, namely the domain wall feature corresponds to stripes in the latter.

In conclusion, we have proposed an experimental way on resonance Fermion superfluid systems with unequal mixtures of two species to achieve the FFLO state. One attractive point to accomplish it is that under a fixed population difference, say 10$\%$ by moving out from the Feshbach resonance point, we can cross the boundary from the uniform Cooper paired state into the FFLO state. This boundary region is most favorable to image the modulated pattern
since the magnetization has a sharp peak at the zero of the pairing field.
The direct imaging can be performed either by species-selective absorption, or by Stern-Gerlach separation.

One of the authors (K.M.) thanks W. Ketterle for encouragement.


\begin{thebibliography}{99}
\bibitem{memo}
In the present work, we restrict ourselves to the non-symmetric system, i.e., $\mathcal{K}_{\uparrow}-\mu _{\uparrow} \neq \mathcal{K}_{\downarrow} - \mu _{\downarrow}$. When $\mathcal{K}_{\uparrow}-\mu _{\uparrow} = \mathcal{K}_{\downarrow} - \mu _{\downarrow}$, the particle-hole symmetry is restored where that the eigenstates ($u_{\bf q},v_{\bf q})^{T}$ for $\varepsilon _{\bf q} > 0$ are equivalent to $(-v^{\ast}_{\bf q},u^{\ast}_{\bf q})^{T}$ for $\varepsilon _{\bf q}<0$. In this case, the usual self-consistent equations in the BCS theory are obtained by the transformation $\varepsilon _{\bf q} \rightarrow - \varepsilon _{\bf q}$, $u_{\bf q} \rightarrow - v^{\ast}_{\bf q}$, and $v_{\bf q} \rightarrow u ^{\ast}_{\bf q}$.
\end{thebibliography}
\end{document}